\documentclass[fleqn,12pt,twoside]{article}
\usepackage{espcrc1}
\usepackage[figuresright]{rotating}
\usepackage{graphicx}
\usepackage{epsfig}

\newcommand{\AmS}{{\protect\the\textfont2
  A\kern-.1667em\lower.5ex\hbox{M}\kern-.125emS}}
\title{Charged Hadron Spectra in PHENIX}
\author{J.Jia\address{Department of Physics and Astronomy, \\
State University of New York at Stony Brook, Stony Brook, NY 11794-3800, USA}
for the PHENIX Collaboration{\thanks{for the full PHENIX Collaboration author list and acknowledgements, see Appendix "Collaborations" of this volume.}}}
\begin{document}
\maketitle

\begin{abstract}
PHENIX has measured transverse momentum spectra of charged particles at mid-rapidity up to 10 GeV/c in Au+Au collisions at $\sqrt{s}_{nn}=200$ GeV. For central collisions, the yield at high $p_{T}$ is significantly suppressed comparing to binary scaled p+p data, and this suppression is stronger than observed at 130 GeV \cite{ppg03,ppg13}. Above 4 GeV/c, the deficit of high momentum particles is almost independent of $p_{T}$.
\end{abstract}

\section{Introduction}
\label{sec:intro}
In collisions at RHIC beam energies, jet fragmentation is expected to dominate the production of high $p_{T}$ particles. According to theoretical calculations \cite{wang1}, jets, which are produced early in the collisions, will suffer significant energy loss via gluon radiation when passing through the hot dense medium. As a consequence, the particle yield at high $p_{T}$ should be significantly suppressed \cite{wang2}. First data from RHIC have revealed such a suppression \cite{ppg03}. Since the amount of energy loss is sensitive to the density and the path length through the medium, the suppression effect should depend on centrality \cite{wang3}. A study of the suppression and its evolution with collision centrality may help to constrain theoretical models and calculations. 
\section{Analysis and corrections}
In this analysis, we used 27.6 million minimum-bias events triggered by the coincidence of 2 Zero-Degree Calorimeters (ZDC) and 2 Beam-Beam Counters (BBC). Event centrality is determined by correlating BBC charge and ZDC energy. Charged particles are tracked in west arm of PHENIX by a Drift Chamber (DC) and 3 Pad Chambers (PC1,PC2,PC3), after passing through an azimuthally symmetric magnetic field. Particle momenta are measured with high resolution ($\rm \delta p/p \simeq 1\%   \oplus 1\% \  p$ (GeV/c)). Particle trajectories are confirmed by a tight track match to the measured position in all three Pad Chambers. Tracks generated close to the DC by decay and photon conversion are only partially deflected in the magnetic field, and may be falsely reconstructed as high $p_{T}$ tracks. However the typical momentum for such tracks is less than 1 GeV/c. As a result of their large multiple scattering and their deflection in the residual magnetic field, most of these tracks fail to correctly match with hits in the outer pad chambers. Those tracks which do match constitute a non-neglegible background above 6 GeV/c. This $p_{T}$ dependent background is measured and subtracted statistically. A conservative estimate of the systematic error of the background subtractions is 10\% at 6 GeV/c and 30\% at 10 GeV/c. 
\indent\par
To correct the raw momentum distributions for acceptance, decay in flight, reconstruction efficiency and momentum resolution, events with single particle are simulated through the GEANT implementation of the PHENIX detector and reconstructed using the same analysis algorithms as applied to the real data. The dependence of the correction on the detector occupancy is studied by embedding single simulated tracks into real events. The correction is a factor 1.4 for the 5\% most central collisions, independent of $p_{T}$.

\section{Results and Discussion}
\begin{figure}[htb]
\vspace{-0.5cm}
\begin{minipage}{.44\linewidth}
  \begin{flushleft}
  \begin{center}
  \epsfig{file=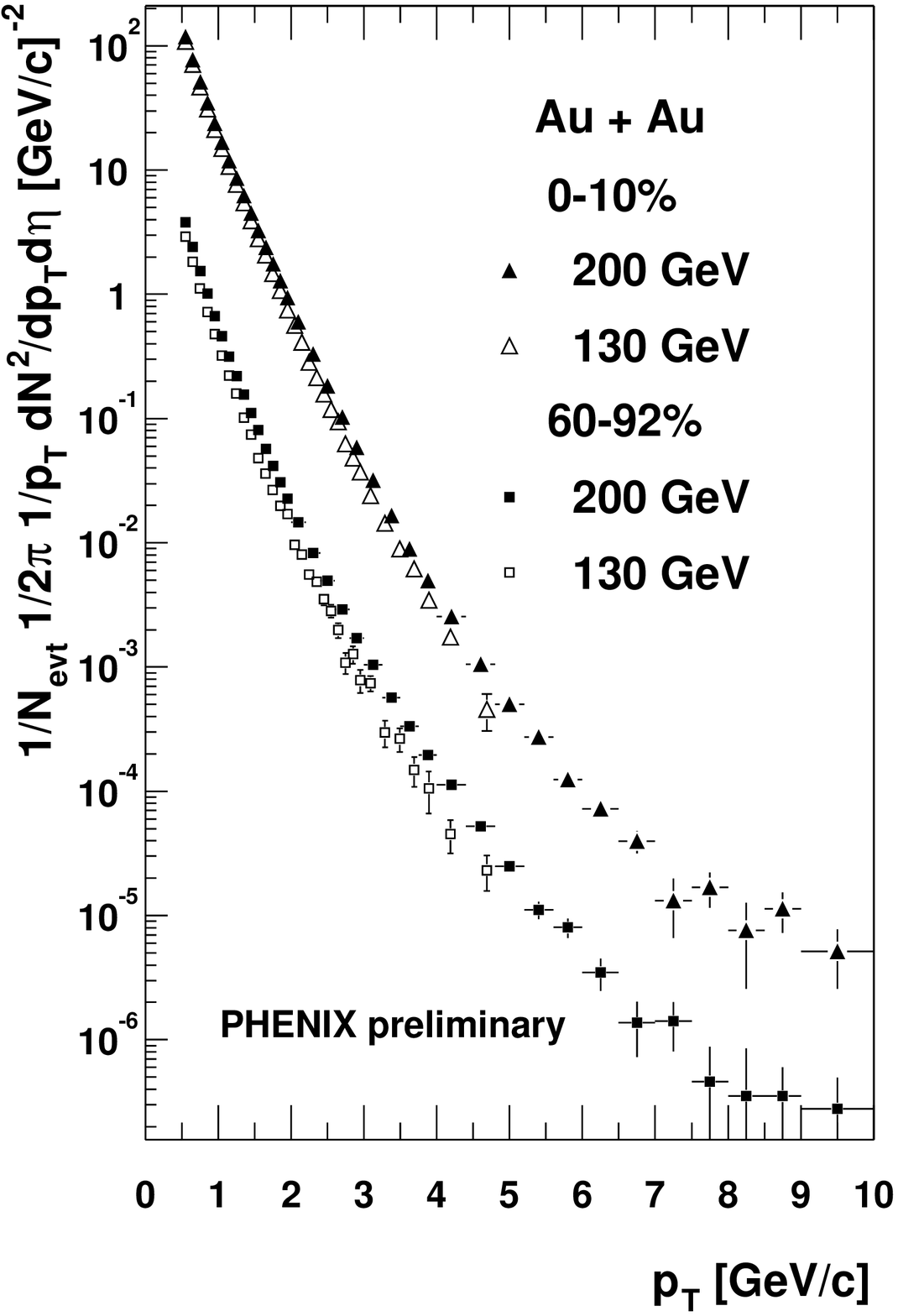,width=1.1\linewidth,height=1.3\linewidth}
 \end{center}
\end{flushleft}
\end{minipage}
\begin{minipage}{.54\linewidth}
  \begin{flushright}	
  \begin{center}
     	\epsfig{file=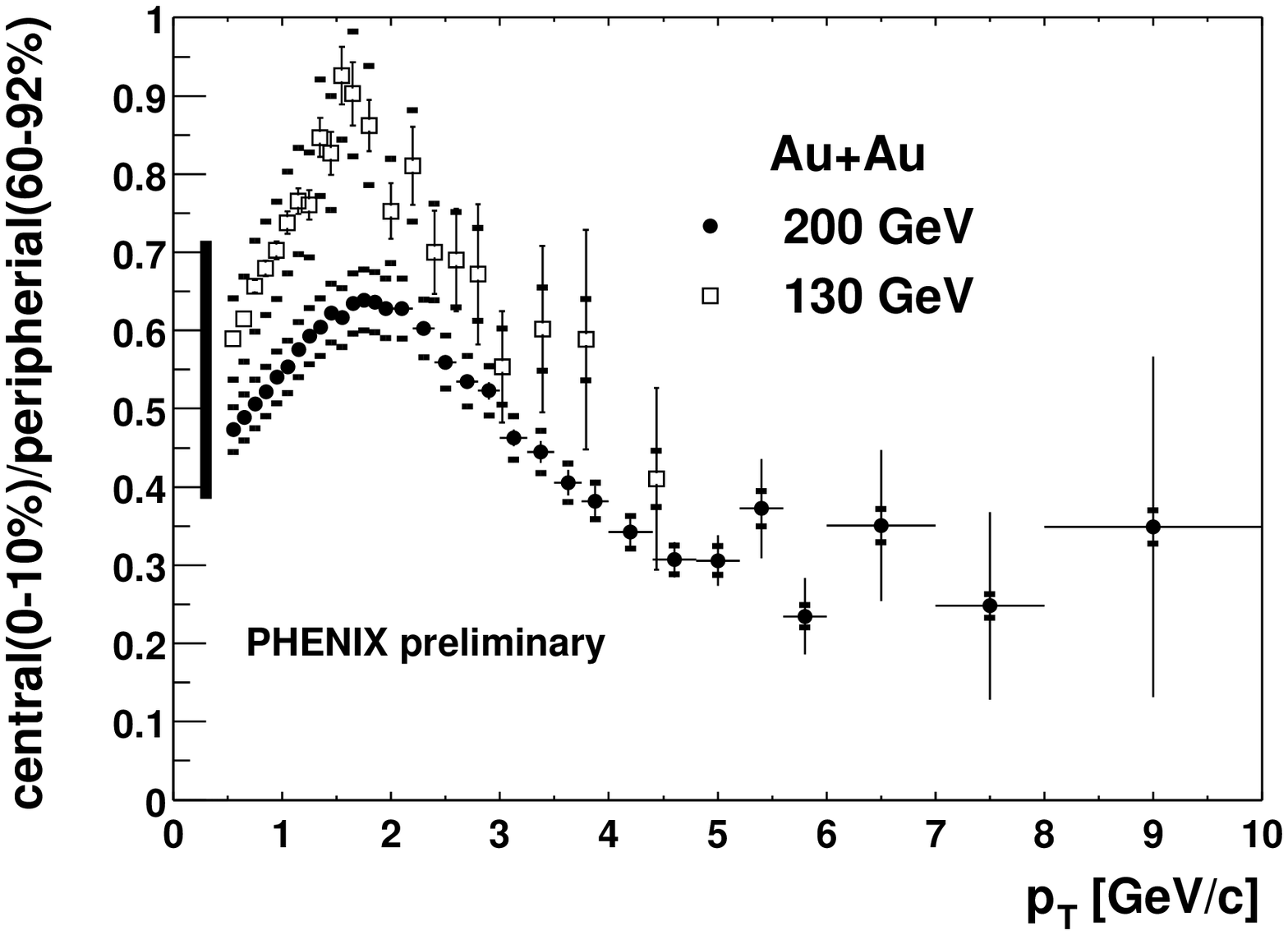,width=1.1\linewidth}
\end{center}
\end{flushright}	
\end{minipage}

\vspace{-1.4cm}

\begin{flushleft}	
\begin{minipage}{.44\linewidth}
\begin{center}
  \caption{{\label{fig:spectra}}\footnotesize
  Charged hadron spectra for central and peripheral collisions for both $\sqrt{s_{nn}} = 200$ GeV and $\sqrt{s_{nn}} = 130$ GeV. The systematic errors are roughly 16\% below 8 GeV/c and 33\% above 8GeV/c, they are not shown for clarity.}
\end{center}
\end{minipage}
\end{flushleft}
\vspace*{-4.7cm}
\begin{flushright}	
  \begin{minipage}{.50\linewidth}
  \begin{center}
     	\caption{{\label{fig:ratio}} \footnotesize
	  Ratio of central to peripheral charged particle spectra for two energies. The solid error bar on each point is the statistical error, the brackets around each point indicate the systematic error, the vertical thick line at the left edge indicates the overall scale uncertainty on the ratio due to the uncertainty on the number of binary collisions.}
  \end{center}
  \end{minipage}
  \end{flushright}	
\vspace*{0cm}
\end{figure}

Invariant charged hadron spectra at mid-rapidity are presented in Fig.~\ref{fig:spectra}
for two different centrality selections ($0-10\%$ and $60-92\%$) and two energies ($\sqrt{s_{nn}} = 200$ GeV and $\sqrt{s_{nn}} = 130$ GeV). The spectra obtained at 200 GeV develop a clear power law shape at high $p_{T}$ and are higher in yield relative to 130 GeV. The increase in yield is consistent with the increase of total multiplicity in PHENIX \cite{phobos,mul1,sasha}. The ratio between central and peripheral collisions, both normalized by the mean number of binary collisions, is plotted in Fig.~\ref{fig:ratio}. The ratio at 200 GeV is somewhat lower than at 130 GeV, possibly reflecting a stronger suppression or a decrease of the proton contribution from 130 to 200 GeV \cite{chujo,saskia}. In general, the ratios for two energies have similar trends, namely an increase up to 2 GeV/c followed by a decrease as a function of $p_{T}$. Within errors, the ratio is approximately constant above 4 GeV/c.

\indent\par
\begin{figure}[htb]
 \hspace{-0.5cm}
  \begin{minipage}{0.48\linewidth}
  \begin{flushleft}
     	\epsfig{file=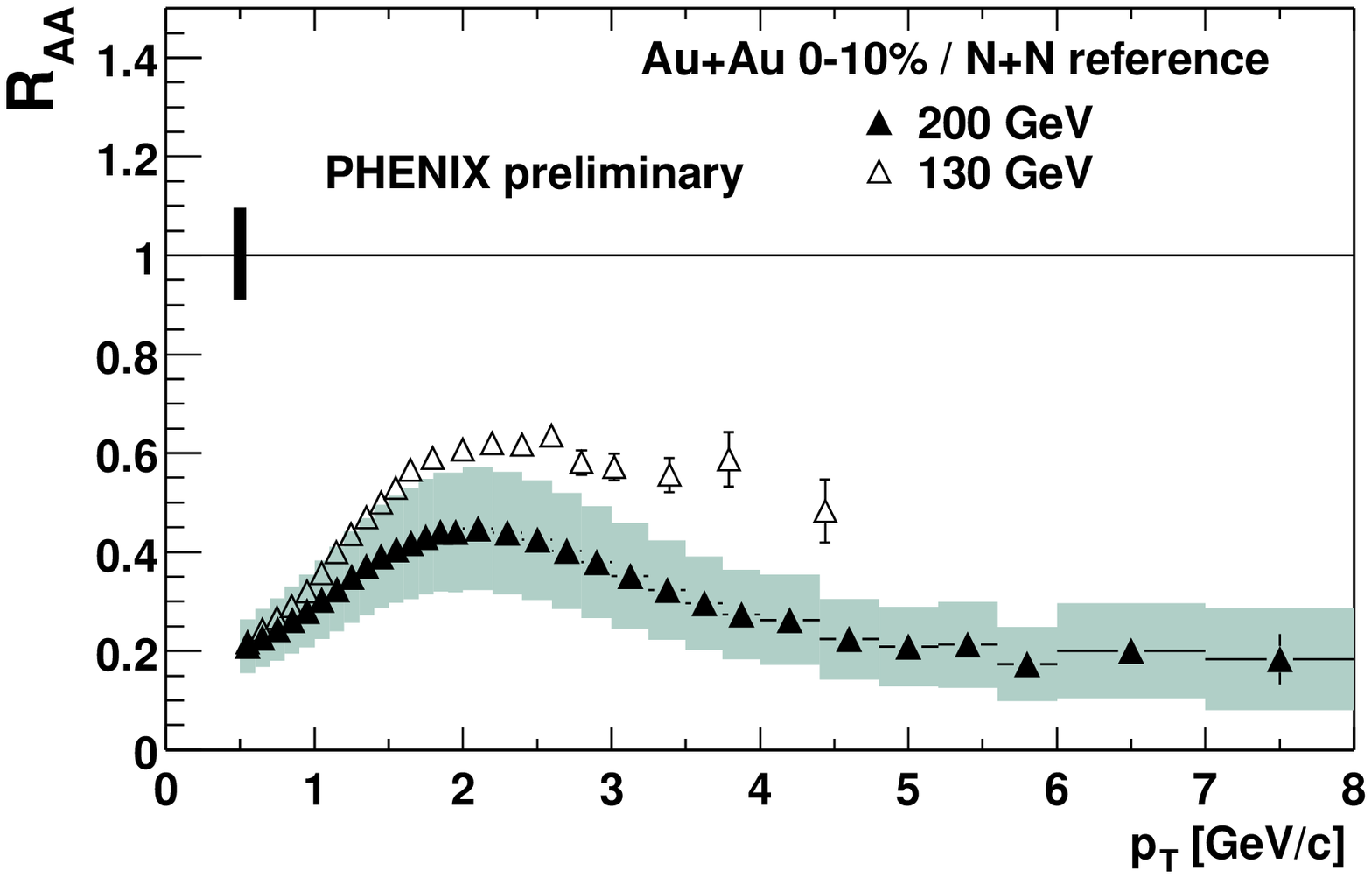,width=1.12\linewidth}
  \end{flushleft}
  \end{minipage}
  \begin{minipage}{0.48\linewidth}
  \begin{flushright}
     	\epsfig{file=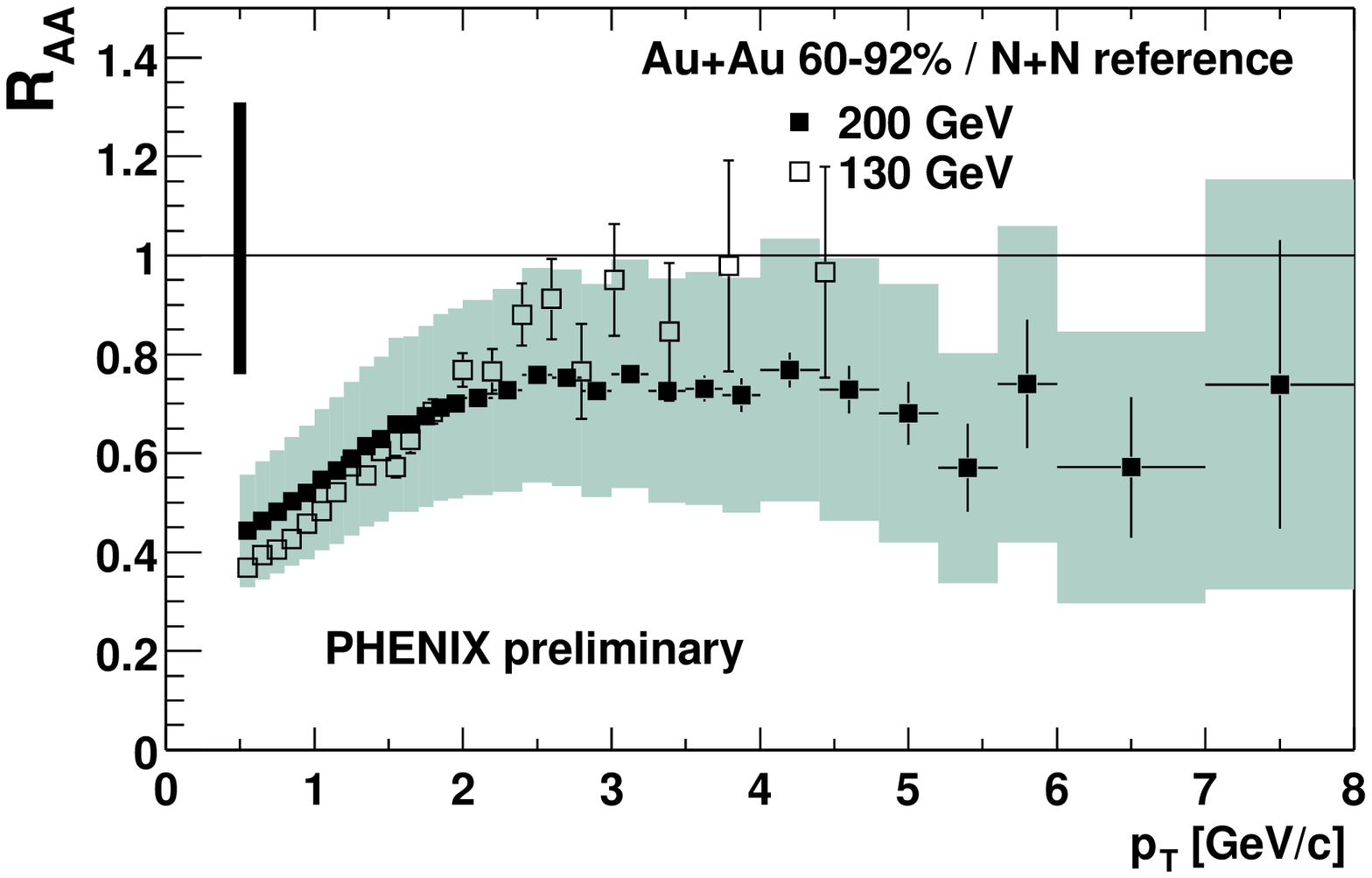,width=1.12\linewidth}
  \end{flushright}
  \end{minipage}

\vspace*{0cm}
 \hspace{1cm}
  \begin{minipage}{0.8\linewidth}
     	\caption{{\label{fig:raa1}} \footnotesize
	  Ratio of Au+Au collisions to the p+p reference \cite{torii} (central collisions on left panel,peripheral collisions on right panel) for both $\sqrt{s_{nn}} = 200$ GeV and $\sqrt{s_{nn}} = 130$ GeV. The shaded band indicates the uncertainty on the p+p reference and systematic error of the spectra. For $\sqrt{s_{nn}} = 130$ GeV, this uncertainty is somewhat smaller. The vertical bar at the left edge indicates the $p_{T}$ independent scale error due to uncertainty on the number of collisions.}
  \end{minipage}
\vspace*{0.5cm} 
\vspace*{0cm} 
  \begin{minipage}{0.49\linewidth}
  \begin{flushleft}
     	\epsfig{file=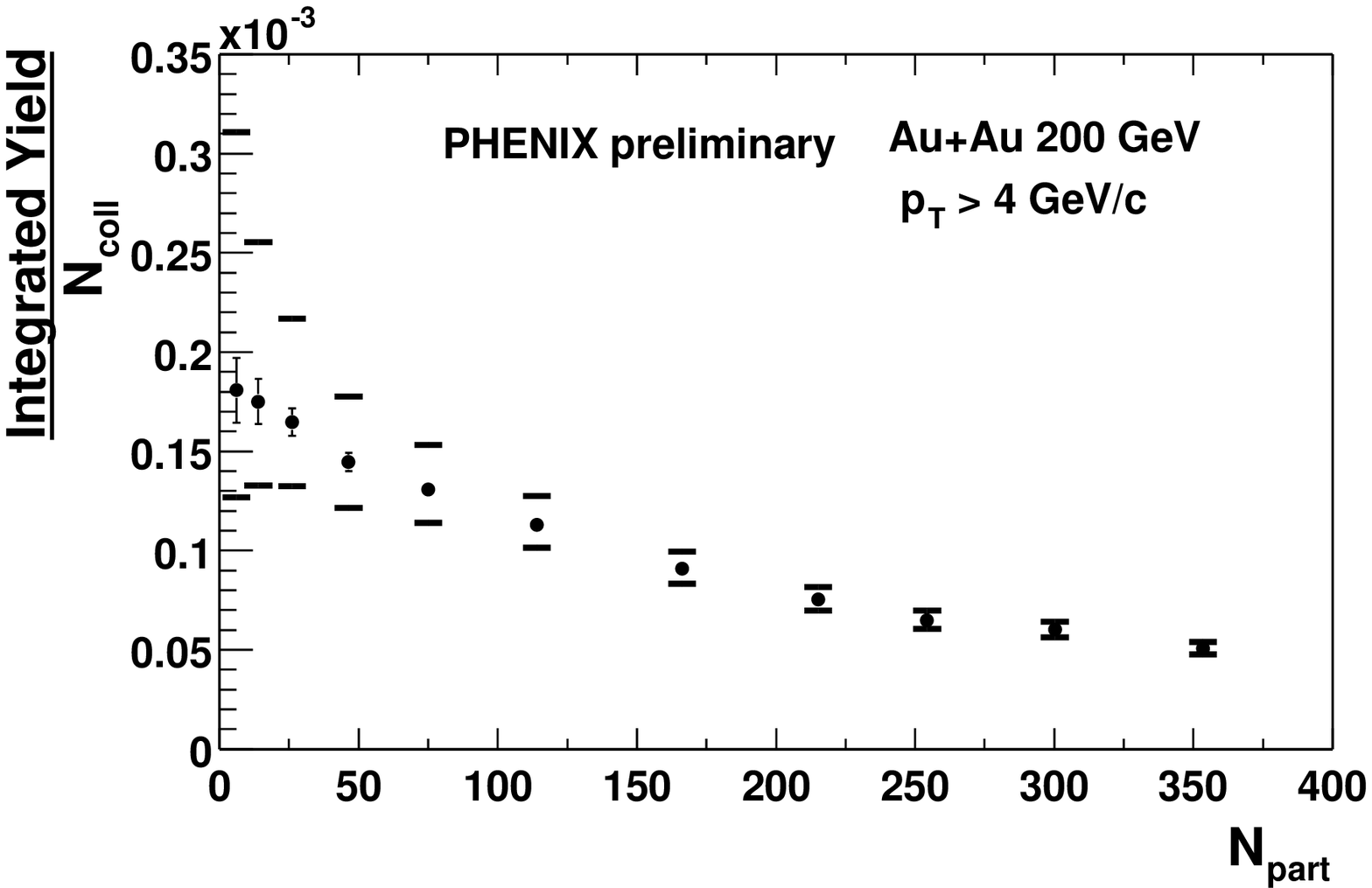,width=1.1\linewidth}
  \end{flushleft}
  \end{minipage}
  \begin{minipage}{0.49\linewidth}
  \begin{flushright}
     	\epsfig{file=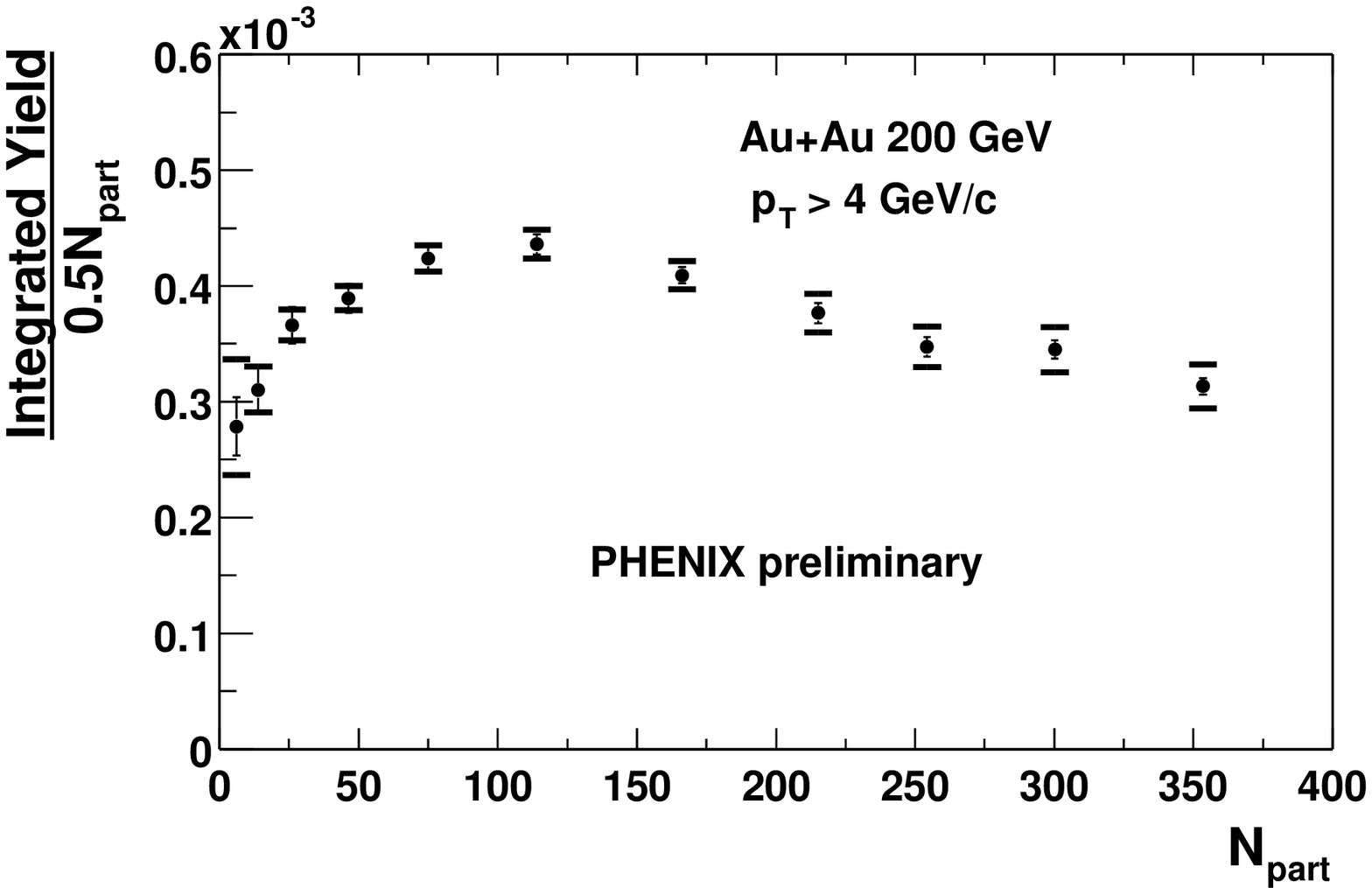,width=1.1\linewidth}
  \end{flushright}
  \end{minipage}

\vspace*{0cm}
 \hspace*{1.5cm}
  \begin{minipage}{0.8\linewidth}
     	\caption{{\label{fig:yield1}} \footnotesize
	  Integrated charged hadron yield above 4 GeV/c nomalized by  $\langle N_{coll} \rangle$ (left) and $\langle N_{part-pair} \rangle$ (right). Brackets around data points indicates the systematic errors due to the uncertainty of $\langle N_{coll} \rangle$ / $\langle N_{part-pair} \rangle$ and the occupancy correction.}
  \end{minipage}
\vspace*{0.5cm} 
\end{figure}

\vspace{-0.5cm}
The suppression effect can be quantified by $R_{AA}$, the {\it nuclear modification factor}, which is defined as

\begin{equation}
R_{AA}(p_\perp) =\frac{1}{\langle N_{binary}\rangle}\frac{dN_{AA}/dyd^2p_\perp}{dN_{pp}/dyd^2p_\perp}
\label{equ:raa}
\end{equation}
The p+p reference spectrum was obtained from a simultaneous fit to 200 GeV UA1 data \cite{ua1} and PHENIX $\pi^{0}$ data scaled by a factor of 1.6 \cite{torii}. The $p_{T}$ dependence of $R_{AA}$ is shown for peripheral and central collisions in Fig.~\ref{fig:raa1}. In peripheral collisions, the $R_{AA}$ values agree within errors over the entire $p_{T}$ range, while in central collisions the $R_{AA}$ at 200 GeV is significantly lower than at 130 GeV over the entire $p_{T}$ range. In addition, at $p_{T} > 4$ GeV/c, both $R_{AA}$'s for 200 GeV, especially the one for cental collisions, are nearly flat, consistent with the central to peripheral ratio shown in Fig.\ref{fig:ratio}.

\vspace{0cm} 

\indent\par
Finally, the centrality dependence of the suppression is shown by Fig.~\ref{fig:yield1}. In left panel, the charged particle yield integrated above 4 GeV/c and normalized to the number of binary collisions is plotted as a function of centrality. We observe a continuous decrease of the yield per-collision with increasing $N_{part}$; the per-collision yield in the most central collisions is a factor of 3.5 smaller than in the most peripheral collisions. For comparision, we show in the right panel of Fig.~\ref{fig:yield1} the integrated yield normalized by the number of participant pairs. The yield per participant pair increases with Npart for Npart $<$ 100 and then decreases again for Npart $>$ 100. This decrease in more central collisions indicats that starting from mid-central ($40-50\%$) collisions, the suppression increases faster than the number of participant pairs.

\def\IJMPA{{Int. J. Mod. Phys.}~{\bf A}}
\def\JPG{{J. Phys}~{\bf G}}
\def\NCA{Nuovo Cimento}
\def\NIM{Nucl. Instrum. Methods}
\def\NIMA{{Nucl. Instrum. Methods}~{\bf A}}
\def\NPA{{Nucl. Phys.}~{\bf A}}
\def\NPB{{Nucl. Phys.}~{\bf B}}
\def\PLB{{Phys. Lett.}~{\bf B}}
\def\PLC{Phys. Repts.\ }
\def\PRL{Phys. Rev. Lett.\ }
\def\PRD{{Phys. Rev.}~{\bf D}}
\def\PRC{{Phys. Rev.}~{\bf C}}
\def\ZPC{{Z. Phys.}~{\bf C}}
\def\EPJC{{Eur.Phys.J.}~{\bf C}}

\end{document}